\documentstyle{article}
\begin{document}

\begin{center} 
{\Large\bf SPECIAL FUNCTIONS AND PATHWAYS\\
FOR PROBLEMS IN ASTROPHYSICS:\\[0.3cm]

AN ESSAY IN HONOR OF A.M. MATHAI}\\[0.3cm]

Hans J. Haubold\\[0.3cm]
Office for Outer Space Affairs, United Nations\\
Vienna International Centre, P.O. Box 500, A-1400 Vienna, Austria\\
WWW: neutrino.aquaphoenix.com\\

and\\

Centre for Mathematical Sciences Pala Campus\\
Arunapuram P.O., Pala, Kerala-686574, India\\
WWW: www.cmsintl.org\\
\end{center}

\noindent
Abstract. The paper provides a review of A.M. Mathai's applications of the theory of special functions, particularly generalized hypergeometric functions, to problems in stellar physics and formation of structure in the Universe and to questions related to reaction, diffusion, and reaction-diffusion models. The essay also highlights Mathai's recent work on entropic, distributional, and differential pathways to basic concepts in statistical mechanics, making use of his earlier research results in information and statistical distribution theory. The results presented in the essay cover a period of time in Mathai's research from 1982 to 2008 and are all related to the thematic area of the gravitationally stabilized solar fusion reactor and fractional reaction-diffusion, taking into account concepts of non-extensive statistical mechanics. The time period referred to above coincides also with Mathai's exceptional contributions to the establishment and operation of the Centre for Mathematical Sciences, India, as well as the holding of the United Nations (UN)/European Space Agency (ESA)/National Aeronautics and Space Administration (NASA) of the United States/ Japanese Aerospace Exploration Agency (JAXA) Workshops on basic space science and the International Heliophysical Year 2007, around the world. Professor Mathai's contributions to the latter, since 1991, are a testimony for his social conscience applied to international scientific activity.  
\clearpage

\section {BIOGRAPHICAL NOTES:\\
Professor Dr. A.M. Mathai}
\noindent
Arakaparampil Mathai Mathai was born in Arakulam in the Idukki district of Kerala, India. He is the eldest son of Aley and Arakaparampil Mathai. He completed his High School education in 1953 with record marks in St. Thomas High School, Palai. He then entered St. Thomas College in Palai, obtained his BSc degree in mathematics in 1957 and went to the University of Kerala, Trivandrum, where he completed his MSc degree in statistics in 1959 with first class, first rank, and a golden medal. He joined St. Thomas College as a lecturer from 1959 to 1961. In 1961, he was awarded a Canadian Commonwealth Scholarship and went to the University of Toronto were he completed an MA and a PhD in  mathematics and mathematical statistics in only three years, the minimum required time. He then joined McGill University in Montreal - one of the most prestigious universities in Canada - where he was Assistant Professor of Mathematics and Statistics, 1964-1968, Associate Professor, 1968-1978, and Full Professor, 1979-2000. Since 2000, he is Emeritus Professor in Mathematics and Statistics at McGill University, Montreal, Canada. He served and is currently serving on editorial boards of a number of journals and he has supervised over 30 Master's students and 5 PhD students. Currently he is the Director of the Centre for Mathematical Sciences, Trivandrum and Pala Campuses, Kerala, India which he is developing into an national, regional, and international centre of excellence for research in the mathematical sciences. The holding of Seven six-week SERC Schools in 1995, 2000, 2005, 2006, 2007, 2008, and 2009 are a testament of highest achievements in teaching and research in the mathematical sciences for the Centre. 

\noindent
A.M. Mathai is a Fellow of the National Academy of Sciences, India, and a Fellow of the Institute of Mathematical Statistics, United States of America. He is an Elected Member of the International Statistical Institute, The Netherlands. He is the Founder of the Canadian Journal of Statistics, Canada, and the Founder of the Statistical Science Association of Canada (now, Statistical Society of Canada). Professor Mathai is a Honorary Member of the International Scientific Organizing Committee of the United Nations/European Space Agency/National Aeronautics and Space Administration of the United States of America/Japanese Aerospace Exploration Agency Workshops on basic space science which have been held, since 1991, in India, Costa Rica, Colombia, Nigeria, Egypt, Sri Lanka, Germany, Honduras, Jordan, France, Mauritius, Argentina, China, United Arab Emirates, India, Japan, Bulgaria, and South Korea. Professor Mathai is author or co-author of fourteen research level books, seven undergraduate-graduate level books, and more than thirty edited and co-edited proceedings. More than 250 published research papers in journals from around the world are showing the breadth of the array of topics that he has tackled as well as the depth, the complexity and the originality of his published results put him in a class all his own. All these publications concern A.M. Mathai's research areas with significant research contributions in mathematical statistics, applied statistics, probability, physics, mathematical physics, astrophysics, information theory, mathematics, and specific problems in biological models, orthogonal polynomials, dispersion theory, axiomatization of basic statistical concepts, pollution problems, and transportation problems. An electronic library, recording all of A.M. Mathai's publications, is under preparation at the United Nations in Vienna.

\noindent
Around the world, A.M. Mathai is known for being very energetic, exceptionally industrious and remarkably gifted. Those individuals who have the good fortune to be close acquaintances of his are also aware of his moral integrity and his social conscience. The latter was particularly beneficial to have him as the advisor in the organization of scientific activities under the umbrella of the United Nations. 

\clearpage

\section {ASTROPHYSICAL THERMONUCLEAR\\
FUNCTIONS}
\noindent
Stars are gravitationally stabilized fusion
reactors changing their chemical composition while transforming
light atomic nuclei into heavy ones. The atomic nuclei are supposed
to be in thermal equilibrium with the ambient plasma (Maxwell-Boltzmann). The majority
of reactions among nuclei leading to a nuclear transformation are
inhibited by the necessity for the charged particles to tunnel
through their mutual Coulomb barrier (Gamow). As theoretical knowledge and
experimental verification of nuclear cross sections increases it
becomes possible to refine analytic representations for nuclear
reaction rates. 
\noindent

\subsection{The Integrals}
\noindent
The standard case of the thermonuclear
function contains a nuclear cross section, the energy
dependent term of a Taylor series, and the Maxwell-Boltzmann distribution function,
\begin{equation}
I_1(z,\nu) \stackrel{def}{=} \int_0^\infty y^\nu
e^{-y}e^{-zy^{-\frac{1}{2}}}dy\label{eq:2.4}
\end{equation}
where $y=E/kT$ and $z=2\pi(\mu/2kT)^{1/2}Z_1Z_2e^2/\hbar$.
Considering dissipative collision processes in the thermonuclear
plasma, cut off of the high energy tail of the Maxwell-Boltzmann
distribution may occur, thus we write,
\begin{equation}
I_2(z,d,\nu) \stackrel{def}{=}  \int_0^d y^\nu
e^{-y}e^{-zy^{-\frac{1}{2}}}dy,\label{eq:2.5}
\end{equation}
where d denotes a certain typically high energy threshold.\par
Accommodating screening effects in the standard thermonuclear
function we have to use nuclear cross section and the
Maxwell-Boltzmann distribution function which
leads to 
\begin{equation}
I_3(z,t,\nu) \stackrel{def}{=} \int_0^\infty y^\nu
e^{-y}e^{-z(y+t)^{-\frac{1}
{2}}}dy,\label{eq:2.6}
\end{equation}
where t is the electron screening parameter. \par
Finally, if due to plasma effects a depletion of the Maxwell-
Boltzmann distribution has to be taken into account, the
thermonuclear function can be written in the following form
\begin{equation}
I_4(z,\delta,b,\nu) \stackrel{def}{=} \int_0^\infty y^\nu
e^{-y}e^{-by^\delta}e^{-zy^{-\frac{1}{2}}}dy, \label{eq:2.7}    
\end{equation}
where the parameter $\delta$ exhibits the enhancement or reduction
of the high-energy tail of the Maxwell-Boltzmann distribution.\par
Mathematical/statistical techniques can be used for
deriving closed-form representations of the above thermonuclear
functions. 

\subsection{Closed-form Representation of the Integrals}
\noindent
For $z>0$ and $\nu\geq 0$, we have
\begin{equation}
I_1(z,\nu)=\int_0^\infty y^\nu e^{-y}e^{-zy^{\frac{1}
{2}}}dy=\pi^{\frac{-1}
{2}}G^{3,0}_{0,3}\left(\frac{z^2}{4}\bigg|_{0,\frac{1}
{2},1+\nu}\right).\label{eq:4.6}
\end{equation}
\noindent
For $z>0,d>0$, and $\nu\geq0$, we have 
\begin{equation}
I_2(z,d,\nu)=\int_0^d y^\nu e^{-y}e^{-zy^{\frac{-1}
{2}}}dy=\frac{d^{1+\nu}}{\pi^{\frac{1}
{2}}}\sum_{r=0}^\infty\frac{(-d)^r}{r!}G^{3,0}_{1,3}\left(\frac{z
^2}
{4d}\bigg|_{\nu+r+1,0,\frac{1}{2}}^{\nu+r+2}\right).\label{eq:4.7}
\end{equation} 
\noindent
We have
\begin{eqnarray}
I_3(z,t,\nu) & \stackrel{def}{=} & e^t\sum_{r=0}^\nu(^\nu
_r)(-t)^{\nu-r}\left[I_1(z,r)-I_2(z,t,r)\right]\label{eq:4.8}
\end{eqnarray}
\noindent
We have
\begin{eqnarray}
I_4(z,\delta,b,\nu)& \stackrel{ref}{=} \sum_{r=0}^\infty\frac{(-b)^r}
{r!}I_1(z,\nu+r\delta).\label{eq:4.9}
\end{eqnarray}
We obtain 
\begin{eqnarray}
I_1(z,\nu) & \sim & 2\left({\pi\over 3}\right)^{1\over
2}\left({z^2\over 4}\right)^{2\nu+1\over
6}e^{-3\left({z^2\over 4}\right)^{1/3}},\label{eq:4.10}\\
I_2(z,d,\nu) & \sim & d^{\nu+1}e^{-d}\left({z^2\over
4d}\right)^{-1/2}e^{-2\left({z^2\over
4d}\right)^{1/2}},\label{eq:4.11}\\
I_3(z,t,\nu) & \sim & 2\left({\pi\over 3}\right)^{1\over
2}e^t\left({z^2\over 4}\right)^{1\over 6}e^{-3 
\left({z^2\over 4}\right)^{1/3}}\left[\left({z^2\over
4}\right)^{1\over
3}-t\right]^\nu,\label{eq:4.12}\\
I_4(z,\delta,b,\nu) &\sim & 2\left({\pi\over 3}\right)^{1\over
2}\left({z^2\over 4}\right)^{2\nu+1\over
6}e^{-3\left({z^2\over 4}\right)^{1/3}}e^{-b\left({z^2\over
4}\right)^{\delta/3}},\label{eq:4.13}
\end{eqnarray}
all as $z\to\infty$.

\subsection {Further Reading}
\noindent
Anderson, W.J., Haubold, H.J., and Mathai, A.M.: 1992, Astrophysical thermonuclear functions, {\sl Astrophysics and Space Science}, {\bf 214}, 49-70.\par
\smallskip
\noindent
Bergstrom, Iguri, S., and Rubinstein, H.: 1999, Constraints on the variation of the fine structure constant from big bang nucleosynthesis, {\sl Physical Review} {\bf D60}, 045005.\par
\smallskip
\noindent
Critchfield, C.L.: 1972, Analytic forms of the thermonuclear function. In: {\sl Cosmology, Fusion, and Other Matters. George Gamow Memorial Volume}, Edited by F. Reines, University of Colorado Press, Colorado, pp. 186-191.\par
\smallskip
\noindent
Davis Jr., R.: 1996, A review of measurements of the solar neutrino flux and their variation, {\sl Nuclear Physics, Proceedings Supplement}, {\bf B48}, 284-298.\par
\smallskip
\noindent
Fowler, W.A.: 1984, Experimental and theoretical nuclear astrophysics: The quest for the origin of the elements, {Reviews of Modern Physics} {\bf 56}, 149-179.\par
\smallskip
\noindent
Haubold, H.J. and Kumar, D.: 2008, Extension of thermonuclear functions through the pathway model including Maxwell-Boltzmann and Tsallis distributions, {\sl Astroparticle Physics}, {\bf 29}, 70-76.\par
\smallskip
\noindent
Mathai, A.M., Haubold, H.J.: 1988, {\sl Modern Problems in Nuclear and Neutrino Astrophysics}, Akademie-Verlag, Berlin.\par
\smallskip
\noindent
Mathai, A.M., Haubold, H.J.: 2002, Review of mathematical techniques applicable in astrophysical reaction rate theory, {\sl Astrophysics and Space Science}, {\bf 282}, 265-280.\par
\smallskip
\noindent
Mathai, A.M., Haubold, H.J.: 2008, {\sl Special Functions for Applied Scientists}, Springer-Verlag, New York.\par 
\smallskip
\noindent
Mathai, A.M., Saxena, R.K.: 1973, {\sl Generalized Hypergeometric Functions with Applications in Statistics and Physical Sciences}, Lecture Notes in Mathematics Vol. 348, Springer-Verlag, Berlin.\par
\smallskip
\noindent
Saxena, R. K.: 1960, {\sl Proceedings of the National Academy of Sciences, India} {\bf 26}, 400-413.\par
\smallskip
\noindent

\section {SOLAR STRUCTURE IN TERMS OF\\
GAUSS' HYPERGEOMETRIC FUNCTION}
\noindent
Hydrostatic equilibrium and energy conservation
determine
the conditions in the gravitationally stabilized solar fusion
reactor. We assume a matter density distribution varying
non-linearly through the central region of the Sun. The analytic
solutions of the differential equations of mass conservation,
hydrostatic equilibrium, and energy conservation, together with
the
equation of state of the perfect gas and a nuclear energy
generation rate $\epsilon=\epsilon_0 \rho^nT^m$, are given in
terms
of Gauss' hypergeometric function. This model for the structure
of
the Sun gives the run of density, mass, pressure, temperature,
and
nuclear energy generation through the central region of the Sun.
Because of the assumption of a matter density distribution, the
conditions of hydrostatic equilibrium and energy conservation are
separated from the mode of energy transport in the Sun.

\subsection{Hydrostatic Equilibrium}
\noindent
In the following we are concerned with the hydrostatic
equilibrium
of the purely
gaseous spherical central region of the Sun generating energy by
nuclear reactions at a certain rate. For this gaseous sphere we
assume that the matter density varies non-linearly from the
center
outward, depending on two parameters $\delta$ and $\gamma$,
\begin{equation}
\rho(x)=\rho_cf_D(x),
\end{equation}
\begin{equation}
f_D(x)=[1-x^\delta]^\gamma,
\end{equation}
where $x$ denotes the dimensionless distance variable,
$x=r/R_\odot,
0\leq x\leq 1,$ $R_\odot$ is the solar radius, $\delta >0,
\gamma>0$
and $\gamma$ is kept a positive integer in the following
considerations. The
choice of the density distribution reveals
immediately that $\rho(x=0)=\rho_c$ is the central density of the
configuration and $\rho(x=1)=0$ is a boundary condition for
hydrostatic equilibrium of the gaseous configuration. For the
range
$0\leq x\leq0.3$ the density distribution can be fit numerically to computed data for solar models by
choosing
$\delta=1.28$ and $\gamma=10$. The choice of restricting $x$ to $x\leq 0.3$
is justified by looking at a Standard Solar Model which shows that $x\leq 0.3$ comprises what
is considered to be the gravitationally stabilized solar fusion
reactor. More precisely, 95\% of the solar luminosity is produced
within the region $x<0.2 (M<0.3M_\odot)$. The half-peak value for
the matter density occurs at $x=0.1$ and the half-peak value for
the temperature occurs at $x=0.25$. The region $x\leq0.3$ is also
the place where the solar neutrino fluxes are generated.
As we are concerned with a spherically symmetrical distribution
of
matter, the mass $M(x)$ within the radius $x$ having the density
distribution is
\begin{equation}
M(x)=M_\odot f_M(x),
\end{equation}
\begin{equation}
f_M(x)=\frac{(\frac{3}{\delta}+1)(\frac{3}{\delta}+2)\cdots
(\frac{3}{\delta}+\gamma)}{\gamma!}x^3\, _2F_1(-\gamma,
\frac{3}{\delta};\frac{3}{\delta}+1;
x^\delta),
\end{equation}
where $M_\odot$ denotes the solar mass and $_2F_1(.)$ is
Gauss' hypergeometric function.
This equation is satisfying
the boundary condition $M(x=0)=0$ and determine the central value
$\rho_c$ of the matter density through the boundary condition
$M(x=1)=M_\odot$, where $\rho_c$ depends then only on $\delta$
and
$\gamma$ of the chosen density distribution.\par
For hydrostatic equilibrium of the gaseous configuration the
internal pressure needs to balance the gravitational attraction.
The pressure distribution follows by integration of the
respective differential equation for hydrostatic equilibrium,
making use of the above density distribution and mass
distribution, that is
\begin{equation}
P(x)=\frac{9}{4\pi}G\frac{M_\odot^2}{R_\odot^4}f_P(x),
\end{equation}
\begin{eqnarray}
f_P(x)&=&
\left[\frac{(\frac{3}{\delta}+1)(\frac{3}{\delta}+2)\ldots(\frac{
3}
{\delta}+\gamma)}{\gamma!}\right]^2\frac{1}{\delta^2}
\sum^\gamma_{m=0}\frac{(-\gamma)_m}{m!
(\frac{3}{\delta}+m)(\frac{2}
{\delta}+m)}\nonumber \\
& & \times
\left[\frac{\gamma!}{(\frac{2}{\delta}+m+1)_\gamma}-x^{\delta
m+2}\,_2F_1(-\gamma,
\frac{2}{\delta}+m;\frac{2}{\delta}+m+1;x^\delta)\right],
\end{eqnarray}
where $G$ is Newton's constant and $_2F_1(.)$ denotes again
Gauss'
hypergeometric function.\par
The Pochhammer symbol
$(\frac{2}{\delta}+m+1)_\gamma=\Gamma(\frac{2}{\delta}+m+1+
\gamma)/\Gamma(\frac{2}{\delta}+m+1)$ often appears in series
expansions
for hypergeometric functions. This equation gives the
value of the pressure $P_c$ at the centre of the gaseous configuration
and satisfies the condition $P(x=1)=0.$.

\subsection{Equation of State}
\noindent
It should be noted that $P(x)$ denotes the total
pressure of the gaseous configuration, that is the sum of the gas
pressure and the radiation pressure (according to
Stefan-Boltzmann's law).
However, the radiation pressure, although the ratio of radiation
pressure to gas pressure increases
towards the center of the Sun, remains negligibly small in
comparison to the gas pressure. Thus, $P(x)$ can be
considered to represent the run of the gas pressure through the
configuration under consideration. Further, the matter density is
so low that at the temperatures involved the material follows the
equation of state of the perfect gas. Therefore, the temperature
distribution throughout the gaseous configuration is given by
\begin{equation}
T(x)=3\frac{\mu}{kN_A}G\frac{M_\odot}{R_\odot}f_T(x),
\end{equation}
\begin{eqnarray}
f_T(x)&=&\left[\frac{(\frac{3}{\delta}+1)(\frac{3}{\delta}+2)\cdots
(\frac{3}{\delta}+\gamma)}{\gamma!}\right]\frac{1}{\delta^2}\frac
{1}{[1-x^
\delta]^\gamma}\sum^\gamma_{m=0}
\frac{(-\gamma)_m}{m!(\frac{3}
{\delta}+m)(\frac{2}{\delta}+m)}\nonumber \\
& &
\times\left[\frac{\gamma!}{(\frac{2}{\delta}+m+1)_\gamma}-x^{\delta
m+2}\, _2F_1(-\gamma,
\frac{2}{\delta}+m;\frac{2}{\delta}+m+1;x^\delta)\right],
\end{eqnarray}
where $k$ is the Boltzmann constant, $N_A$ Avogadro's number,
$\mu$
the mean molecular weight, and $_2F_1(.)$ Gauss' hypergeometric
function. The equations for $T(x)$ reveal the central
temperature for
$T(x=0)=T_c$ and satisfy the boundary condition $T(x=1)=0.$ Since
the gas in the central region of the Sun can be treated as
completely ionized, the mean molecular weight $\mu$ is given by
$\mu=(2X+\frac{3}{4}Y+\frac{1}{2}Z)^{-1},$ where $X, Y, Z$ are
relative abundances by mass of hydrogen, helium, and heavy
elements, respectively, and $X+Y+Z=1.$

\subsection{Nuclear Energy Generation Rate}
\noindent
Hydrostatic equilibrium and energy conservation are determining
the
physical conditions in the central part of the Sun. In the
preceeding Sections, the run of density, mass, pressure, and
temperature have been given for a gaseous configuration in
hydrostatic equilibrium based on the equation of state of the
perfect gas. In the following, a
representation for
the nuclear energy generation rate,
\begin{equation}
\epsilon(\rho, T)=\epsilon_0 \rho^nT^m,
\end{equation}
will be sought which takes into account the above given
distributions of density and temperature and which can be used to
integrate the differential equation of energy conservation
throughout the gaseous configuration. In the above
equation, $n$ denotes the density exponent, $m$ the
temperature
exponent, and $\epsilon_0$ a positive constant determined by the
specific reactions for the generation of nuclear energy. 

\subsection{Total Nuclear Energy Generation}
\noindent
The total net rate of nuclear energy generation is equal to the
luminosity of the Sun, that means the generation of energy by
nuclear reactions in the central part of the Sun has to
continually
replenish that energy radiated away at the surface. If $L(x)$
denotes the outflow of energy across the spherical surface at
distance $x$ from the center, then in equilibrium the average
energy production at distance $x$ is
\begin{equation}
L(x)=4\pi R^3_\odot\int_0^x dt t^2\rho (t)\epsilon(t),
\end{equation}
where $\rho(x)$ and $\epsilon(x)$ are given above.

\subsection {Further Reading}
\noindent
Chandrasekhar, S.: 1939/1957, {\sl An Introduction to the Study of Stellar Structure}, Dover Publications Inc., New York.\par
\smallskip
\noindent
Clayton, D.D.: 1986, Solar structure without computers, American Journal of Physics, {\bf 54}, 354-362.\par
\smallskip
\noindent
Haubold, H.J., Mathai, A.M.: 1995, Solar structure in terms of Gauss'hypergeometric function, {\sl Astrophysics and Space Science}, {\bf 228}, 77-86.\par
\smallskip
\noindent
Haubold, H.J.: 1995, An analytic solar model: Physical principles and mathematical structures, {\sl International Journal of Mathematical and Statistical Sciences}, {\bf 4}, 31-41.\par
\smallskip
\noindent
Haubold, H.J., Mathai, A.M.: 2001, Sun, in {\sl Encyclopedia of Planetary Sciences}, eds. J.H. Shirley and R.W. Fairbridge, Chapman and Hall, New York, pp. 786-794.\par
\smallskip
\noindent
Mathai, A.M. and Haubold, H.J.: 2008, {\sl Special Functions for Applied Physicists}, Springer, New York.\par
\smallskip
\noindent
Mathai, A.M., Saxena, R.K., and Haubold, H.J.: 2009, {\sl The H-Function: Theory and Applications}, Springer, New York.\par
\smallskip
\noindent
Stein, R.F.: 1966, in R.F. Stein and A.G.W. Cameron (eds.), {\sl Stellar Evolution}, Plenum Press, New York.\par

\noindent
\section {THE FRACTIONAL KINETIC EQUATION}\par
\noindent
The subject of this remark is to derive the solution of the fractional kinetic equations. The results are obtained in a compact form containing the Mittag-Leffler function, which naturally occurs whenever one is dealing with fractional integral equations.

\subsection{Kinetic Equation and Mittag-Leffler Function}
\noindent
In terms of Pochammer's symbol
$$(\alpha)_n=\left\{^{1,n=0}_{\alpha(\alpha+1)\ldots(\alpha+n-1),n\in N}\right.$$
we can express the binomial series as
\begin{equation}
(1-x)^{-\alpha}=\sum^\infty_{r=0}\frac{(\alpha)_rx^r}{r!}.
\end{equation}
The Mittag-Leffler function is defined by
\begin{equation}
E_\alpha(x):=\sum^\infty_{n=0}\frac{z^n}{\Gamma(\alpha n+1)},
\end{equation}
This function was defined and studied by Mittag-Leffler.
We note that this function is a direct generalization of an exponential function, since
$$E_1(z):=exp(z).$$
It also includes the error functions and other related functions, we have
\begin{equation}
E_{1/2}(\pm z^{1/2})=e^z[1+erf(\pm z^{1/2})]=e^z erfc(\mp z^{1/2}),
\end{equation}
where \begin{equation}
erf(z):=\frac{2}{\pi^{1/2}}\int^z_0e^{-u^2}du, erfc(z):= 1-erf(z), z\in C.
\end{equation}
The equation
\begin{equation}
E_{\alpha,\beta}(z):=\sum^\infty_{n=0}\frac{z^n}{\Gamma(\alpha n+\beta)}
\end{equation}
gives a generalization of the Mittag-Leffler function.
When $\beta = 1$, the generalization reduces to the original.
Both functions are entire functions of order $1/\alpha$ and type 1.
The Laplace transform of $E_{\alpha, \beta}(z)$ follows from the integral
\begin{equation}
\int^\infty_0e^{-pt}t^{\beta-1}E_{\alpha, \beta}(\lambda at^\alpha)dt=p^{-\beta}(1-ap^{-\alpha})^{-1},
\end{equation}
where $Re(p)>|a|^{1/\alpha}, Re(\beta)>0$, which can be established by means of the Laplace integral
\begin{equation}
\int^\infty_0e^{-pt}t^{\rho-1}dt=\Gamma(\rho)/p^\rho,
\end{equation}
where $Re(p)>0, Re(\rho)>0$.
The Riemann-Liouville operator of fractional integration is defined as
\begin{equation}
_aD_t^{-\nu}f(t)=\frac{1}{\Gamma(\nu)}\int^t_a  f(u)(t-u)^{\nu-1}du, \nu>0,
\end{equation}
with $_aD_t^0 f(t)=f(t)$.
By integrating the standard kinetic equation
\begin{equation}
\frac{d}{dt}N_i(t)=-c_iN_i(t), (c_i>0),
\end{equation}
it is derived that
\begin{equation}
N_i(t)-N_0=-c_i\;\;_0D_t^{-1}N_i(t),
\end{equation}
where $_0D_t^{-1}$ is the standard Riemann integral operator. The number density of species $i, N_i=N_i(t)$, is a function of time and $N_i(t=0)=N_0$ is the number density of species $i$ at time $t=0$. By dropping the index $i$, the solution of its generalized form
\begin{equation}
N(t)-N_0=-c^\nu\;\;_0D_t^{-\nu}N(t),
\end{equation}
is obtained as 
\begin{equation}
N(t)=N_0\sum^\infty_{k=0}\frac{(-1)^k(ct)^{\nu k}}{\Gamma(\nu k+1)},
\end{equation}
By virtue of the original definition of the Mittag-Leffler function we can rewrite this solution in terms of the Mittag-Leffler function in a compact form as
\begin{equation}
N(t)=N_0 E_\nu(-c^\nu t^\nu), \nu>0.
\end{equation}

\subsection {Conclusions}
\noindent
Solutions are given in terms of the ordinary Mittag-Leffler function and their generalization, which can also be represented as FOX's H-functions. The ordinary and generalized Mittag-Leffler functions interpolate between a purely exponential law and power-like behavior of phenomena governed by ordinary kinetic equations and their fractional counterparts.

\subsection {Further Reading}
\noindent
Grafiychuk, V., Datsko, B., and Meleshko, V.: 2006, Mathematical modeling of pattern formation in sub- and superdiffusive reaction-diffusion systems, arXiv:nlin.AO/0611005 v3.\par
\smallskip
\noindent
Haubold, H.J. and Mathai, A.M.: 2000, The fractional kinetic equation and thermonuclear functions, {\it Astrophysics and Space Science,} {\bf 327}, 53-63.\par
\smallskip
\noindent
Mathai, A.M. and Haubold, H.J.: 2008, {\sl Special Functions for Applied Physicists}, Springer, New York.\par
\smallskip
\noindent
Mathai, A.M., Saxena, R.K., and Haubold, H.J.: 2009, {\sl The H-Function: Theory and Applications}, Springer, New York.\par
\smallskip
\noindent
Oldham, K.B. and Spanier, J.: 1974, {\sl The Fractional Calculus: Theory and Applications of Differentiation and Integration to Arbitrary Order}, Academic Press, New York.\par
\smallskip
\noindent
Saxena, R.K., Mathai, A.M., Haubold, H.J.: 2002, On fractional kinetic equations, {\sl Astrophysics and Space Science}, {\bf 282}, 281-287.\par
\smallskip
\noindent
Saxena, R.K., Mathai, A.M., Haubold, H.J.: 2004, Unified fractional kinetic equation and a fractional diffusion equation, {\sl Astrophysics and Space Science}, {\bf 290}, 299-310.\par
\smallskip
\noindent
Saxena, R.K., Mathai, A.M., and Haubold, H.J.: 2006, Fractional reaction-diffusion equations, {\it Astrophysics and Space Science,} {\bf 305}, 289-296.\par
\smallskip
\noindent
Srivastava, H.M. and Saxena, R.K.: 2001, Operators of fractional integration and their applications, {\it Applied Mathematics and Computation,} {\bf 118}, 1-52.\par
\smallskip
\noindent
Tsallis, C.: 2004, What should a statistical mechanics satisfy to reflect nature?, {\sl Physica D}, {\bf 193}, 3-34.\par
\smallskip
\noindent
Turing, A.M.: 1952, The chemical basis of morphogenesis, {\it Philosophical Transactions of the Royal Society of London,} {\bf B237}, 37-72.\par
\smallskip
\noindent 

\section {GENERALIZED MEASURE OF ENTROPY, MATHAI PATHWAYS, AND TSALLIS\\
STATISTICS}\par
\noindent
When deriving or fitting models for data from physical experiments
very often the practice is to take a member from a parametric family of distributions. 
But it is often 
found that the model requires a distribution with a more specific tail than the
ones available from the parametric family, or a 
situation of right tail cut-off. The model reveals that  the underlying 
distribution is in between two parametric families of distributions. In order 
to create a distributional pathway for proceeding from one functional form to another a pathway 
parameter is introduced and a pathway model is created. This 
model enables one to go from a generalized type-1 beta model to a generalized 
type-2 beta model to a generalized gamma model when the variable is restricted 
to be positive. More families are available when the variable is allowed to vary
over the real line. Corresponding multitudes of families are available when the 
variable is in the complex domain. Originally, Mathai deals mainly with rectangular 
matrix-variate 
distributions in the real case and the scalar case is the topic of discussion in the current
remark because the scalar case is connected to many problems in physics. Similarly, through a generalized measure of entropy, an entropic pathway can be devised that leads to different families of entropic functionals that produce families of distributions by applying the maximum entropy principle.
\subsection{The Pathway Model in the Real Scalar Case}
\noindent
For the real 
scalar case the pathway model for positive random variables is the following:
\begin{equation}
f(x)=c\;x^{\gamma -1}[1-a(1-\alpha)x^{\delta}]^ {\frac{\beta}{1-\alpha}},\;\; x>0,
\end{equation}
$a>0,\delta >0, \beta\geq 0,1-a(1-\alpha)x^{\delta}>0,\gamma >0$ where $c$ is the 
normalizing constant and $\alpha$ is the pathway parameter.  The normalizing 
 constant in this case is the following:
\begin{eqnarray}
c&=&\frac{\delta [a(1-\alpha)]^{\frac{\gamma}{\delta}}
\Gamma\left(\frac{\gamma}{\delta}+\frac{\beta}{1-\alpha}+1\right)} 
{\Gamma\left(\frac{\gamma}{\delta}\right)\Gamma\left(\frac{\beta}{1-\alpha}
+1\right)},\;\;\mbox{for}\; \alpha<1 \\
&=& \frac{\delta [a(\alpha-1)]^{\frac{\gamma}{\delta}}\Gamma\left(\frac{\beta}{\alpha-1}\right)}
{\Gamma\left(\frac{\gamma}{\delta}\right)\Gamma\left(\frac{\beta}{\alpha-1}
-\frac{\gamma}{\delta}\right)},\;\;\mbox{for}\;\;\frac{1}{\alpha-1}
-\frac{\gamma}{\delta}>0,\;\;\alpha>1 \\
&=&\frac{\delta\;(a\beta)^{\frac{\gamma}{\delta}}}{\Gamma\left(\frac{\gamma}
{\delta}\right)},\;\mbox{for}\;\;\alpha\rightarrow 1.
\end{eqnarray}
\noindent

\subsection{Pathway Model from a Generalized Entropy Measure}
\noindent
A generalized entropy measure of order $\alpha$
is a generalization of Shannon entropy and it is  a variant of the 
generalized entropy of order 
$\alpha$. In the discrete case the measure is the 
 following: 
Consider a multinomial population $P=(p_1,\ldots, p_k)$,$\;p_i\geq 0$,$\; i=1,\ldots, k$, 
$p_1+\ldots +p_k=1$. Define the function
\begin{equation}
M_{k,\alpha}(P)=\frac{\sum_{i=1}^k p_i^{2-\alpha}-1}
{\alpha-1},\;\alpha\neq 1,\;\;-\infty<\alpha<2 
\end{equation}
\begin{equation}
\lim_{\alpha\rightarrow 1}M_{k,\alpha}(P) =-\sum_{i=1}^k p_i\ln p_i
=S_k(P)
\end{equation}
by using L'Hospital's rule. In this notation $0\ln 0$ is taken as zero when 
any $p_i=0$. Thus it is a generalization of Shannon entropy $S_k(P)$. Note that it is a variant of Havrda-Charv\'at entropy 
$H_{k,\alpha}(P)$ and Tsallis entropy $T_{k,\alpha}(P)$ where
\begin{equation}
H_{k,\alpha}(P)=\frac{\sum_{i=1}^kp_i^\alpha-1}{2^{1-\alpha}-1},\;\alpha\neq 1,\;\alpha>0
\end{equation}
and
\begin{equation}
T_{k,\alpha}(P)=\frac{\sum_{1=1}^kp_i^\alpha-1}{1-\alpha},\;\;\alpha\neq 1,\;\;\alpha>0.
\end{equation}
We will introduce another measure associated with the above and parallel to 
R\'enyi entropy $R_{k,\alpha}$ in the following form:
\begin{equation}
M_{k,\alpha}^*(P)=\frac{\ln\left(\sum_{i=1}^k p_i^{2-\alpha}\right)}{\alpha-1},\;\;\alpha\neq 1,-\infty<\alpha<2.
\end{equation}
R\'enyi entropy is given by
\begin{equation}
R_{k,\alpha}(P)=\frac{\ln\left(\sum_{i=1}^k p_i^\alpha \right)}{1-\alpha},\;\; \alpha\neq 1,\;\alpha>0.
\end{equation}
\par 
\noindent
\subsubsection{Continuous analogue}
\noindent
The continuous analogue to the generalized entropy measure is the following:
\begin{eqnarray}
M_\alpha(f)&=& \frac{\int_{-\infty}^\infty[f(x)]^{2-\alpha}dx-1}
{\alpha-1}\\
&=&\frac{\int_{-\infty}^{\infty}[f(x)]^{1-\alpha}f(x)dx-1}
{\alpha-1}=\frac{E[f(x)]^{1-\alpha}-1}{\alpha-1},\;\;\alpha\neq 1,\;\alpha <2,\nonumber
\end{eqnarray}
where $E[\cdot]$ denotes the expected value of $[\cdot]$. 
Note that when $\alpha =1,\\
E[f(x)]^{1-\alpha}=E[f(x)]^0=1.$\\
It is easy to see that the generalized entropy measure is connected to Kerridge's
``inaccuracy" measure. The generalized inaccuracy measure is 
$E[q(x)]^{1-\alpha}$ where the experimenter has assigned $q(x)$ for the true
density $f(x)$, where $q(x)$ could be an estimate of $f(x)$ or $q(x)$ could
be coming from observations. Through disturbance or distortion if the 
experimenter assigns $[f(x)]^{1-\alpha}$ for $[q(x)]^{1-\alpha}$ then
the inaccuracy measure is $M_{\alpha}(f)$. 
\subsubsection{Distributions with maximum generalized entropy}
\noindent
Among all densities, which one will give a maximum value for $M_{\alpha}(f)$? 
Consider all possible functions $f(x)$ such that $f(x)\geq 0$ for all $x$, 
$f(x)=0$ outside $(a,b)$, $a<b$, $f(a)$ is the same for all such $f(x)$, 
$f(b)$ is the same for all such $f$, 
 $\int_{a}^{b}f(x)dx<\infty$.  Let $f(x)$ be a continuous function of $x$ with continuous 
derivatives in $(a,b)$. Let us maximize $\int_{a}^{b}[f(x)]^{2-\alpha}dx$ 
for fixed $\alpha$ and over all functional $f$, under the conditions that the 
following two moment-like expressions be fixed quantities:
\begin{equation}
\int_a^b x^{(\gamma-1)(1-\alpha)}f(x)dx=\mbox{given, and}
\int_a^b x^{(\gamma-1)(1-\alpha)+\delta}f(x) dx=\mbox{given}
\end{equation}
for fixed $\gamma>0$ and $\delta>0$. Consider
\begin{equation}
U=[f(x)]^{2-\alpha}-\lambda_1 x^{(\gamma-1)(1-\alpha)}f(x)+\lambda_2 x^{(\gamma-1)(1-\alpha)+\delta}f(x),\;\alpha<2,\;\alpha\neq 1
\end{equation}
where $\lambda_1$ and $\lambda_2$ are Lagrangian multipliers. Then the Euler equation is the following:
\begin{eqnarray}
\frac{\partial U}{\partial f}= 0 &\Rightarrow& (2-\alpha)[f(x)]^{1-\alpha}-\lambda_1 x^{(\gamma-1)(1-\alpha)}+\lambda_2x^{(\gamma-1)(1-\alpha)+\delta}=0\nonumber\\
&\Rightarrow& [f(x)]^{1-\alpha}=\frac{\lambda_1}{(2-\alpha)} x^{(\gamma-1)(1-\alpha)}[1-\frac{\lambda_2}{\lambda_1}x^{\delta}]\\
&\Rightarrow& f(x)=c\;x^{\gamma-1}[1-\eta(1-\alpha)x^{\delta}]^{\frac {1}{1-\alpha}}
\end{eqnarray}
where $\lambda_1/\lambda_2$ is written as $\eta(1-\alpha)$ with $\eta>0$ 
such that $1-\eta (1-\alpha)x^{\delta}>0$ since $f(x)$ is assumed to be 
non-negative. By using the conditions above we can determine $c$ and $\eta$.
 When the range of $x$ for which $f(x)$ is nonzero is $(0,\infty)$ and when 
$c$ is a normalizing constant, then $f(x)$ is the pathway model of Mathai 
in the scalar case where $\alpha$ is the pathway parameter. When 
$\gamma=1,\delta=1$ in $f(x)$ then $f(x)$ produces the power law. The form above for 
various values of $\lambda_1$ and $\lambda_2$ can produce all the four forms
$$\alpha_1x^{\gamma-1}[1-\beta_1(1-\alpha)x^{\delta}]^{-\frac{1}{1-\alpha}},\;
\alpha_2x^{\gamma-1}[1-\beta_2(1-\alpha)x^{\delta}]^{\frac{1}{1-\alpha}}\mbox{for}\alpha<1$$
and
$$\alpha_3 x^{\gamma-1}[1+\beta_3(\alpha-1)x^{\delta}]^{-\frac{1}{\alpha-1}},\;\alpha_4x^{\gamma-1}[1+\beta_4(\alpha-1)x^{\delta}]^{\frac{1}{\alpha-1}} \mbox{for}\alpha>1$$
with $\alpha_i,\beta_i>0,i=1,2,3,4$. But out of these, the second and the third
 forms can produce densities in $(0,\infty)$. The first and fourth will not be 
converging.

\subsection {Further Reading}\par
\noindent
Gell-Mann, M. and Tsallis, C. (Eds.): 2004, {\it Nonextensive Entropy: Interdisciplinary Applications}. Oxford University Press, New York.\par
\smallskip
\noindent
Mathai, A.M. and Haubold, H.J.: 2008, {\sl Special Functions for Applied Physicists}, Springer, New York.\par
\smallskip
\noindent
Mathai, A.M., Saxena, R.K., and Haubold, H.J.: 2009, {\sl The H-Function: Theory and Applications}, Springer, New York.\par
\smallskip
\noindent
Mathai, A.M.: 2005, A pathway to matrix-variate gamma and normal densities. {\it Linear Algebra and Its Applications}, {\bf 396}, 317-328.\par
\smallskip
\noindent
Mathai, A.M.and Haubold, H.J.: 2007, Pathway model, superstatistics, Tsallis statistics, and a generalized measure of entropy, {\sl Physica A}, {\bf 375}, 110-122.\par
\smallskip
\noindent
Mathai, A.M.and Haubold, H.J.: 2008, Pathway parameter and thermonuclear functions, {\sl Physica A}, {\bf 387}, 2462-2470.\par
\smallskip
\noindent
Mathai, A.M.and Haubold, H.J.: 2008, On generalized distributions and pathways, {\sl Physics Letters A}, {\bf 372}, 2109-2113.\par
\smallskip
\noindent
Mathai, A.M. and Rathie, P.N.: 1975, {\it Basic Concepts in Information Theory and Statistics: Axiomatic Foundations and Applications}, Wiley Halsted, New York and Wiley Eastern, New Delhi.\par
\smallskip
\noindent
Nicolis, G. and Prigogine, I.: 1977, {\it Self-organization in Non-equilibrium Systems: From Dissipative Structures to Order Through Fluctuations}, John Wiley and Sons, New York.\par
\smallskip
\noindent
Tsallis, C.: 2009, {\sl Introduction to Nonextensive Statistical Mechanics - Approaching a Complex World}, Springer, New York.\par

\noindent
\section {GRAVITATIONAL INSTABILITY IN A\\
MULTICOMPONENT COSMOLOGICAL\\
MEDIUM}\par
\noindent
The paper presents results for deriving
closed-form analytic solutions of the
non-relativistic linear perturbation equations, which govern the
evolution of inhomogeneities in a homogeneous spatially flat
multicomponent cosmological model. Mathematical methods to derive
computable
forms of
the perturbations are outlined.\par
\noindent

\subsection{Introduction}\par
\noindent
These closed-form solutions are valid for irregularities
on
scales smaller than the horizon. They may be used for analytical
interpolation between known expressions for short and long
wavelength perturbations. We present the physical
parameters relevant to the problem and the fundamental
differential
equations for arbitrary polytropic index $\gamma_i$ and a
radiation
or matter dominated Universe, respectively. Solutions of this
fundamental equation are provided for relevant parameter sets including the polytropic index
$\gamma_i$ and the expansion law index $\eta$. To catalogue these
solutions in closed-form, the theory of Meijer's
G-
function will be employed.\par 
\noindent

\subsection{Some Parameters of Physical Significance}\par
\noindent
The growth of density inhomogeneities can begin as soon as the
Universe is matter-dominated. Baryonic
inhomogeneities cannot begin to grow until after decoupling
because
until then, baryons are tightly coupled to the photons. After
decoupling, when the Universe is matter-dominated and baryons are
free of the pressure support provided by photons, density
inhomogeneities in the baryons and any other matter components
can
grow. Actually the time of matter-radiation equality is the
initial
epoch for structure formation. In order to fill in the details of
structure formation one needs "initial data" for that epoch. The
initial data required include, among others, (1) the total amount
of non-relativistic matter in the Universe, quantified by
$\Omega_0$, and (2) the composition of the Universe, as
quantified
by the fraction of critical density, $\Omega_i=\rho_i/\rho_c$,
contributed by various components of primordial density
perturbations (i= baryons, Wimps, relativistic particles, etc.).
Here the critical density $\rho_c$ is the total matter density
of the Einstein-de Sitter Universe.
Speculations about the earliest history of the Universe have
provided hints as to the appropriate initial data: $\Omega_0=1$
from inflation; $0.015\leq \Omega_B\leq0.15$ and
$\Omega_{WIMP}\sim0.9$ from inflation, primordial
nucleosynthesis,
and dynamical arguments. In the following we assume that
$\sum \Omega_i=1, \Omega_i=const.$. The cosmological
medium is
considered to be unbounded and that there may be a uniform
background
of relativistic matter. It is further assumed that the matter is
only slightly perturbed from the background cosmological model.
This assumption may give a good description of the behavior of
matter on large scales even when there may be strongly nonlinear
clustering on small scales (primordial objects). Also assumed is
that matter can be approximated as an ideal fluid with pressure a
function of density alone. It consists of $i$ components having
the
densities $\rho_i$ and velocities of sound
$\beta_i^2=dP_i/d\rho_i\propto \rho^{\gamma_i-1}$, when an
equation
of state
\begin{equation}
P_i\propto\rho^{\gamma_i}
\end{equation}
has been taken into account. The $i$ components of the medium are
interrelated through Newton`s field equation $\Delta\phi=4\pi
G\sum\rho_i$, containing the combined density $\sum
\rho_i$ of all
components. Superimposed upon an initially homogeneous and
stationary mass distribution shall be a small perturbation,
represented by a sum of plane waves
\begin{equation}
\delta=\frac{\delta\rho_i}{\rho}=\delta_i(t)e^{ikx},
\end{equation}
where $k=2\pi a/\lambda$ defines the wave number $k$ and
$\lambda$
is the proper wavelength. In the linear approach the system of
second order differential equations describing the evolution of
the
perturbation in the non-relativistic component $i$ is
\begin{equation}
\frac{d^2\delta_i}{dt^2}+2(\frac{\dot{a}}{a})\frac{d\delta_i}{dt}
+k^
2\beta_i^2\delta_i=4\pi G\sum_{j=1}^m\rho_j\delta_j, i=1,...,m,
\end{equation}
where for all perturbations the same wave number k is used. This
equation is valid for all sub-horizon-sized perturbations in any
non-relativistic species, so long as the usual Friedmann equation
for the expansion rate is used. Although the following
investigation of closed-form solutions of the fundamental
differential equation governing the evolution of inhomogeneities
in
a multicomponent cosmological medium is quite general, it will be
presented within the context of the inflationary scenario.
Therefore in the following we consider an Einstein-de Sitter
Universe with zero cosmological constant. After inflation the
Friedmann equation for the cosmological evolution reduces to
\begin{equation}
(\frac{\dot{a}}{a})^2=H^2=\frac{8\pi G}{3}\rho.
\end{equation}
The continuity equation gives $\rho\propto a^{-3}$ and 
one has $a\propto t^{2/3}$ and $\rho_i=\Omega_i/(6\pi Gt^2)$.
During
the expansion the Hubble parameter changes as $H=\eta t^{-1}$,
where the expansion law index is
$\eta=\frac{1}{2}$ in the radiation-dominated epoch and
$\eta=\frac{2}{3}$ in the matter-dominated epoch respectively.
The
wave number
is
proportional to $a^{-1}$ so that $k^2\beta_i^2=k_i^2t^{2(1-\eta-
\gamma_i)}$, where now the constants $k_i$ come from
both the wave
vector and the velocity of sound. If $k_i$ = 0 the sound velocity
$\beta_i$ equals zero, in which case the adiabatic index
$\gamma_i$
loses its sense. Defining the parameter $\alpha_i=2(2-\eta-
\gamma_i)$ that absorbs the adiabatic index as well as the type
of
expansion law we can write:
\begin{equation}
t^2\ddot{\delta}_i(t)+2\eta
t\dot{\delta}_i(t)+k_i^2t^{\alpha_i}\delta_i=\frac{2}{3}\sum^m_
{j=1}
\Omega_j\delta_j(t),\;\;\;\;\;i=1,\ldots,m,
\end{equation}
and dots denote derivatives with respect to time. We introduce a
time operator $\Delta = t\mbox{d}/\mbox{d}t$ and change the
dependent variable, the density perturbation $\delta_i(t),$
dealing
with the function $\Phi_i$ instead of $\delta_i$ by setting
$\delta_i(t) = t^\alpha\Phi_i(t).$ The equation for the
function $\Phi_i$ is then given by
\[\Delta^2\Phi_i
+ b_i\Phi_i = \frac{2}{3}\sum_{j=1}^m \Omega_j\Phi_j,\] where
\begin{equation}
\Phi_i=t^{-\alpha}\delta_i, \; b_i=k_i^2t^{\alpha_i}-\alpha^2, \;
\alpha_i=2(2-\eta-\gamma_i), \; \alpha=-\left(\frac{2\eta-
1}{2}\right).
\end{equation}
Observing that $\Sigma\Omega_j =1$ and operating on both sides of
the differential equation for the $\Phi_i$ by $\Delta^2$ we
obtain
the fundamental equation
\begin{equation}
\Delta^4\Phi_i+\Delta^2(b_i\Phi_i)-
\frac{2}{3}(\Delta^2\Phi_i+b_i\Phi_i)=-\frac{2}{3}\sum_{j=1}^m
b_j\Omega_j\Phi_j, \;\;\;i=1,\ldots m.
\end{equation}
As indicated above for $\eta$ the values $\frac{2}{3}$ and
$\frac{1}{2}$
are significant and in what follows it will be assumed that
$2\geq
\gamma_i \geq \frac{2}{3}$. We have chosen the range of values of
$\gamma_i$ for physical as well as mathematical reasons as will
be
evident later in the analysis.
For some of these parameter values we will consider a
multicomponent medium. Consider the special case
\begin{eqnarray}
b_1 & = & k_1^2t^{\alpha_1}-\frac{(2\eta-1)^2}{4}, \nonumber \\
b_2 & = & b_3=\cdots = b_m=b=k^2t^{\alpha}-\frac{(2\eta-1)^2}{4}
\end{eqnarray}
of the fundamental equation. 
Let $k=0$. For this case the parameters are the following:
\[b_1^*, b_2^*=\pm\left\{ \frac{(2\eta-1)^2}{4\alpha_1^2}\right\}
^{\frac{1}{2}},
b_3^*,b_4^*=\pm\left\{ \frac{1}{\alpha_1^2}\left[\frac{(2\eta-
1)^2}{4}+\frac{2}{3}\right]\right\}^\frac{1}{2},\]
\begin{equation}
a_1^*, a_2^*=-1\pm\left\{ \frac{1}{\alpha_1^2}\left[\frac{(2\eta-
1)^2}{4}+\frac{2}{3}-\frac{2}{3}\Omega_1\right]\right\}
^{\frac{1}
{2}}.
\end{equation}

\subsection{Solution when the Parameters Differ by Integers}\par
\noindent
The general solution are for finite values of t and for the
cases that the $b_j^*$'s do not differ by integers. At two points, namely $(\eta=\frac{2}{3},
\gamma_i=\frac{5}{3})$ and $(\eta=\frac{2}{3}, \gamma_i=1)$ one
has
$b_1^*-b_3^*=-1$ and $b_4^*-b_2^*=-1.$\par
\noindent

\subsection{Modifications of the $G_j's$ for the Cases
$(\eta=\frac{2}{3},\gamma_i=1)$
and $(\eta=\frac{2}{3},
\gamma_i=\frac{5}{3})$}\par
\noindent
Consider, for $x=k_1^2 t^{\alpha_1}/ \alpha_1^2,
\alpha_1\neq 0,$
\begin{eqnarray}
G_1 & = & G_{2,4}^{1,2}\left(x\mid^{a_1^*+1, a_2^*+1}_{b_1^*,
b_2^*, b_3^*, b_4^*}\right) \nonumber \\
& = & \frac{1}{2\pi i}\int_L\frac{\Gamma(\frac{1}{4}+s)\Gamma(-
a_1^*-s)\Gamma(-a_2^*-s)x^{-
s}ds}{\Gamma(\frac{5}{4}-s)\Gamma(-\frac{1}{4}-
s)\Gamma(\frac{9}{4}-s)}.
\end{eqnarray}
Note that a zero coming from $\Gamma(-\frac{1}{4}-s)$ at $s=-
\frac{1}{4}$ coincides with a pole coming from
$\Gamma(\frac{1}{4}+s)$ at $s=-\frac{1}{4}$. This can be removed
by
rewriting as follows:
\begin{equation}
G_1=-\frac{1}{2\pi i}\int_L\frac{\Gamma(\frac{5}{4}+s)\Gamma(-
a_1^*-s)\Gamma(-a_2^*-s)x^{-s}ds}{\Gamma(\frac{5}{4}-
s)\Gamma(\frac{3}{4}-s)\Gamma(\frac{9}{4}-s)}.
\end{equation}
Evaluating this expression as the sum of the residues at the poles of
$\Gamma(\frac{5}{4}+s)$ one has
\begin{eqnarray}
G_1 & = &
-x^{\frac{5}{4}}\frac{\Gamma(-a_1^*+\frac{5}{4})\Gamma(-
a_2^*+\frac{5}{4})}{\Gamma(\frac{10}{4})\Gamma(2)\Gamma(\frac{14}
{
4})} \nonumber \\
& & \times\; _2F_3(-a_1^*+\frac{5}{4}, -a_2^*+\frac{5}{4};
\frac{10}{4}, 2, \frac{14}{4}; -x).
\end{eqnarray}
Evaluating $G_4$ also the same way one has the following:
\begin{eqnarray}
G_4 & = & G_{2,4}^{1,2}\left(x\mid^{a_1^*+1, a_2^*+1}_{b_4^*,
b_1^*, b_2^*, b_3^*}\right) \nonumber \\
& = & \frac{1}{2\pi i}\int_L\frac{\Gamma(-\frac{5}{4}+s)\Gamma(-
a_1^*-s)\Gamma(-a_2^*-s)}{\Gamma(\frac{3}{4}-s)\Gamma(\frac{5}{4}
-
s)\Gamma(-\frac{1}{4}-s)}x^{-s}ds \nonumber \\
& = & -\frac{1}{2\pi i}\int_L\frac{\Gamma(-\frac{1}{4}+s)\Gamma(-
a_1^*-s)\Gamma(-a_2^*-s)}{\Gamma(\frac{3}{4}-s)\Gamma(\frac{9}{4}
-
s)\Gamma(-\frac{1}{4}-s)}x^{-s}ds \nonumber \\
& = & -x^{-
\frac{1}{4}}\frac{\Gamma(-\frac{1}{4}-a_1^*)\Gamma(-
\frac{1}{4}-a_2^*)}{\Gamma(\frac{1}{2})\Gamma(2)\Gamma(-
\frac{1}{2})} \nonumber \\
& & \times\; _2F_3(-\frac{1}{4}-a_1^*, -\frac{1}{4}-a_2^*;
\frac{1}{2},2,-\frac{1}{2};-x).
\end{eqnarray}
Thus the complete solution in this case is of the form
\begin{equation}
\Phi_1=c_1G_1+c_2G_2+c_3G_3+c_4G_4,
\end{equation}
where $c_1, c_2, c_3, c_4$ are arbitrary constants.\par
\noindent

\subsection{Conclusion}\par
\noindent
We have presented closed-form solutions of the non-relativistic
linear perturbation equations which govern the evolution of
inhomogeneities in a spatially flat multicomponent cosmological
medium. The general solutions are catalogued according to the
polytropic index $\gamma_i$ and the expansion law index $\eta$ of
the multicomponent medium . All general solutions are expressed
in
terms of Meijer's G-function and their Mellin-Barnes integral
representation. The proper use of this function simplifies the
derivation of solutions of the linear perturbation equations and
opens ways for its numerical computation.

\subsection {Further Reading}
\noindent
Gailis, R.M. and Frankel, N.E.: 2006, Two-component cosmological fluids with gravitational instabilities, {\it Journal of Mathematical Physics,}  {\bf 47}, 062505.\par
\smallskip
\noindent
Gailis, R.M. and Frankel, N.E.: 2006, Short wavelength analysis of the evolution of perturbations in a two-component cosmological fluid, {\it Journal of Mathematical Physics}, {\bf 47} 062506.\par
\smallskip
\noindent
Gottloeber, S., Haubold, H.J., Muecket, J.P., and Mueller, V.: 1988, {\sl Early Evolution of the Universe and Formation of Structure}, Akademie-Verlag, Berlin.\par
\smallskip
\noindent
Haubold, H.J. and Mathai, A.M.: 1994, Computational aspects of the gravitational instability problem for a multicomponent cosmological medium, {\sl Astrophysics and Space Science}, {\bf 214}, 139-149.\par
\smallskip
\noindent
Haubold, H.J. and Mathai, A.M.: 1998, Structure of the Universe, in {\sl Encyclopedia of Applied Physics}, ed. G.L. Trigg, {\bf Vol. 25}, Wiley-VCH Verlag GmbH, Weinheim - New York, 47-81.\par
\smallskip
\noindent
Mathai, A.M.: 1989, On a system of differential equations connected with the gravitational instability in a multicomponent medium in Newtonian cosmology, {\it Studies in Applied Mathematics,} {\bf 80},75-93.\par
\smallskip
\noindent
Mathai, A.M., Saxena, R.K.: 1973, {\it Generalized Hypergeometric Functions with Applications in Statistics and Physical Sciences}. Lecture Notes in Mathematics, Vol. 348, Springer-Verlag, Berlin.\par
\smallskip
\noindent
Mathai, A.M. and Haubold, H.J.: 2008, {\sl Special Functions for Applied Physicists}, Springer, New York.\par
\smallskip
\noindent
Mathai, A.M., Saxena, R.K., and Haubold, H.J.: 2009, {\sl The H-Function: Theory and Applications}, Springer, New York.\par
\smallskip
\noindent
Mather, J.C. et al.: 1992, Early results from the Cosmic Background Explorer (COBE), {\it AIP Conference Proceedings,} {\bf Vol. 245}, American Institute of Physics, New York, 266-276.\par
\smallskip
\noindent
Peebles, P.J.E.: 1980, {\it The Large-Scale Structure of the Universe}, Princeton University Press, Princeton.\par

\section {Developing Basic Space Science World-Wide}
\noindent
Since 1991, Professor Dr. A.M. Mathai was a member of the International Scientific Organizing Committee of all UN/ESA/NASA/JAXA workshops and provided his expertise in the mathematical and statistical sciences to the development of the scientific programmes of these workshops. From 1991 to 2004 the workshops focused on basic space science. From 2005 to 2009 the workshops were dedicated to the International Heliophysical Year 2007. Beginning in 2010, the workshops will develop the International Space Weather Initiative.

\subsection {Further Reading}
\noindent
Wamsteker, W., Albrecht, R., and Haubold, H.J. (Eds.): 2004, {\sl Developing Space Science World-Wide: A Decade of UN/ESA Workshops}, Kluwer Academic Publishers, Dordrecht Boston London.\par
\smallskip
\noindent
Thompson, B.J., Gopalswamy, N., Davila, J.M., and Haubold, H.J. (Eds.): 2009, {\sl Putting the "I" in IHY: The United Nations Report for the International Heliophysical Year 2007}, Springer, New York.\par
\smallskip 
\noindent
Briand, C., Antonucci, E., Gopalswamy, N., Eichhorn, G., Sakurai, T., Mathai, A.M., and Haubold, H.J. (Eds.): 2009, {\sl International Heliophysical Year 2007, Second European General Assembly, Italy, and Third UN/ESA/NASA Workshop, Japan}. {\it Earth, Moon, and Planets,} {\bf Vol. 104}, 1-360.\par

\end{document}